# Theoretical analysis and propositions for "ontology citation"

Biswanath Dutta
DRTC, Indian Statistical Institute
8th Mile Mysore Road
R. V. College Post, Bangalore
bisu@drtc.isibang.ac.in

**Abstract**. Ontology citation, the practice of referring the ontology in a similar fashion the scientific community routinely follows in providing the bibliographic references to other scholarly works, has not received enough attention it supposed to. Interestingly, so far none of the existing standard citation styles (e.g., APA, CMOS, and IEEE) have included ontology as a citable information source in the list of citable information sources such as journal article, book, website, etc. Also, not much work can be found in the literature on this topic though there are various issues and aspects of it that demand a thorough study. For instance, what to cite? Is it the publication that describes the ontology, or the ontology itself? The citation format, style, illustration of motivations of ontology citation, the citation principles, ontology impact factor, citation analysis, and so forth. In this work, we primarily analyse the current ontology citation practices and the related issues. We illustrate the various motivations and the basic principles of ontology citation. We also propose a template for referring the source of ontologies.

## 1. Introduction

A citation, a reference to the source of an information used in a scholarly work, is considered as a way to give credit to the individuals' intellectual creativity and effort (Cronin, 1984). It is largely used as an important medium for measuring the impact of a scholarly work and a scholar. In the recent time, the emphasis on citation has even increased to combat the maniac of plagiarism (University of Pittsburgh, 2018). Usually, the content and style of a citation vary depending on the type of information sources (e.g., books, journals, websites) and the disciplines (e.g., social sciences, humanities, sciences) respectively, although there is a considerable overlap (Wikipedia, n.d.). Some of the examples of popular citation styles in social sciences are American Psychological Association, American Political Science Association, American Anthropological Association, in humanities Chicago Manual Of Style, Modern Language Association, Harvard referencing (or, author-date system), and in sciences American Chemical Society, American Institute of Physics, American Mathematical Society, and Institute of Electrical and Electronics Engineers style. In the recent times, we are observing an increasing emphasis on another type of source citation i.e. data citation. Data citation has become a "*[…] fundamental for considering data as first-class research objects with the same relevance and centrality of traditional scientific products* (Silvello, 2018)". Numerous works can be found in the literature exploring the numerous aspects of data citation (Silvello, 2018, Borgman, 2015).

Besides the traditional scientific publications and the data, in the recent times, we are observing an enormous growth of another type of resources "ontology" often referred as an intelligent knowledge artifact (Dutta, 2017) is used for various knowledge engineering tasks (Dutta, Toulet, Emonet, and Jonquet, 2017). It is a kind of vocabulary (can be both domain specific and domain independent) where the terms and their relationships are formally and explicitly represented (Dutta, 2017; Dutta, Chatterjee, and Madalli, 2017). Note that by ontology, we not only mean the OWL structure that respects all the conditions to be referred as a fully formalized ontology, but also every ontology like structure that formalizes some knowledge (e.g., thesaurus, taxonomy, metadata vocabulary). From the scientific and social use perspectives, the significance of an ontology is no less than the traditional publications and data. Also, from the perspective of its production cost, it is quite expensive (Dutta, Nandini and Shahi, 2015), attracts a great amount of intellectual activity. So, similar to the case of publications and data, it becomes our duty, the users of ontology, to acknowledge and give the due credit to the ontology creators by citing the work whenever and wherever is used. The "ontology citation" (*the practice of referring the ontology in a similar fashion the scientific community routinely follows in providing the bibliographic references to other scholarly works*) will not only enforce the professional ethics but will also motivate the ontology creators to produce quality ontology and publish them. Interestingly, unlike the case of traditional publication citations and data citations, we have not come across much literature on ontology citation. There are a few suggestions found in the ontology specific sites given by the individual ontology creators on how to cite their ontologies (details provided in section 2). It is worthy to mention here that so far even none of the existing standard citation styles (e.g., APA, CMOS, and IEEE) have included ontology as a citable information source in the list of citable information sources such as journal article, book, website, etc. We believe there is a need to give a similar importance to the ontology citation as given to the traditional publication and data citations. There are several aspects of ontology citations that needs to be exploited and analysed, for instance, what would be the style and content of an ontology citation? What to cite? The ontology itself or the publication(s) describing the ontology? Often we find the practice of citing an ontology by referring the ontology *URI* either as a footnote or endnote in a publication. The question is whether this practice is useful enough to achieve the goal of citation and its related aspects? Also, how to refer the ontologies that are being (re)used to build a new ontology? How to analyse the citations? How to calculate the impact factor of an ontology, its creator/ institution? Will the impact factor to be calculated in the same manner calculated for traditional publications? In this work, some of these issues have been further analysed and attempted to provide the possible solutions. The main contributions of this paper are: illustrate the motivations for ontology citation; proposes a set of principles for ontology citation, and also proposes a template for referring the source of ontologies.

## 2. Ontology citation: the current practices and the issues

As expressed above, we have not come across any scientific publication on ontology citation except a few suggestions from the major ontology publishers on how to cite the ontologies *in a publication*. By processing those suggestions, we have found two kinds of practices: (i) the recommendation for citing the specific publication(s) and (ii) the recommendation for citing the ontology using its URI. In the first case, the ontology publisher(s) provides a list of recommended publications to be cited to refer to an ontology. Usually, the recommendations are found in the ontology Website. For instance, in the case of Gene ontology (Gene ontology consortium, n.d.), the site suggests to cite two publications, one is the original publication on GO published in 2000 and the other one is the recent publication in 2017. Similarly, the SUMO ontology site (Pease, 2018) recommends a primary paper and a book to cite SUMO ontology. In the second case, the emphasis is on to cite the ontology URI. For instance, obofoundry.org suggests (OBO foundry, 2018) to cite the ontology URI besides the selected publications, if any. Both of these current practices have some limitations as follows.

Concerning the first recommendation i.e. citing a publication about an ontology in another publication has a direct implication in measuring the ontology impact (i.e. the ontology usage). Our question is how will we be able to make sure that citing a publication means citing an ontology? Because usually a publication about an ontology not only provides the details of the ontology but also provides the general theory and various other aspects of ontology and ontology development. So, when we cite a publication, it is not always necessarily mean that the ontology has been actually used and referred. But it may also possible that the other theoretical aspects of that publication have been referred. So, someone has to manually go through the text of the publication where the ontology related publication(s) has been cited to confirm the actual usage of an ontology. Further, it is not that for all the published ontologies, there are published articles available. A large number of ontologies are produced and registered or deposited in various online ontology libraries (Dutta, Nandini and Shahi, 2015; Naskar and Dutta, 2016), but they do not have any scientific publications. So, the recommendation for citing the publication(s) on ontologies is not always a viable option. Concerning the second recommendation i.e., citing an ontology using its URI involves various issues. The primary issue is URI does not reveal the details of the ontology creators, curators, title, version information, publication year, etc. Hence, one of the motivations of ontology citation i.e. to give due credit to the creators and producers of an ontology remain unaddressed (see section 3 for more on ontology citation motivations).

In the context of the above recommendations for ontology citation and the various issues related to those recommendations, we advocate *citing the actual source of ontology instead of citing the publication(s) describing the ontology*. However, to cite the source of an ontology, there are few obvious questions that we need to answer. The primary questions that can be raised are: what would be the content of the citation referring to the actual source of an ontology? Where and how an ontology to be cited? How to cite an ontology within a publication and within an ontology? What citation style should we follow? In this work, we analyse and provide solutions to some of these questions in the succeeding section.

**3. Why ontology citation? The motivation**

There are several reasons for ontology citations. Some of the most prominent reasons are:

*Ontology attribution*: ontology creation involves a great amount of scientific and intellectual effort. It is important that we give the due credit to the creators of an ontology. The attribution will not only motivate the people to produce, publish and maintain ontologies but also will enable the scientific community to identify the creators and curators of ontologies.

*Ontology discovery*: ontology citation will enable the identification and selection, and provide access to an ontology.

*Ontology impact*: the direct ontology citations (i.e. the citation to an ontology and not the publication that is based on that ontology) will allow us to assess the impact and quality of an ontology based on its actual usage metrics. Because in the case of indirect citations (i.e. the citation to a publication that is based on an ontology), it would be hard to ascertain the reason for citing the publication. For example, as illustrated above, the publication on an ontology may deal in general with the various ontological aspects and related issues. So, when we cite a publication in another publication, it becomes difficult to know whether the citation was given to the ontology or to cite some other ideas embedded in that article. We have to read and manually interpret the reason for citation.

*Ontology visibility*: the citation to an ontology will also increase the ontology visibility. In many of the cases, we see the publications on ontology but we never get to see the actual ontology anywhere on the Web. The reason could be because ontology publication possesses certain technological challenges which may discourage the creators to put an additional effort in publishing the ontologies on the Web. Some of the possible technological challenges are assigning the persistent URI to an ontology, finding a permanent storage space, version control, maintenance, etc. The incentive to the creators in terms of ontology citations may motivate them in publishing the ontologies overcoming the possible technological and technical barriers.

*Ontology sharing*: ontology citation will motivate the authors and organizations to curate and share ontologies. This will inculcate and further spread the culture of ontology sharing.

*Ontology reuse and production*: ontology citation will increase the (re)use and production of new ontologies by providing access to the details of an ontology, for example, the creators, the title, revision information, etc.

*Tracing of ontology history and use*: the ontology citation will enable the quantification of ontology usage in a systematic and consistent way. The ontology citation will also yield information about contributors throughout the development, the usage pattern, and overall will indicate the *quality and value* of an ontology.

*Ontology networking*: ontology citation will enable to find the correlation and networking between the ontologies and also between the ontologists (the creator of an ontology (Naskar and Dutta, 2016)).

## 4. Ontology citation principles

We propose the following *five* ontology citation principles. These principles can be seen from the perspective of solutions to some of the issues discussed in section 2.

*Uniformity*: prepare the citations uniformly, so that the citation data becomes machine processable. The software tools can automatically detect, identify and manipulate the citation data.

*Evidence*: in any scientific publications, the source of an ontology should be cited whenever and wherever it is being used. The reference should be added in the standard reference list of the publication.

*Standard referencing system*: Follow the standard referencing systems. Include the ontology reference in the standard reference list of a publication, and then cite it to the relevant places in the body of the text of the publication.

*Provide complete information*: Provide all the necessary information (and not a mere link) whenever possible and applicable for citing an ontology. For the content of an ontology reference, see section 5.

*Mutual citation*: both the ontology and the publication that is based on that ontology should refer to each other. The reference to an ontology should be added in the standard reference list of a publication as shown in figure 1. On the other hand, the reference to a publication (the publication that is describing the ontology) should be added within the ontology. For citing the publication(s) within an ontology, the *ontology annotation properties* can be exploited. For example, figure 2 shows the reference for the publication on PAV ontology (Ciccarese and Soiland-Reyes, 2014) added in the ontology header of PAV owl file. For adding the reference, we have used the Dublin Core element dcterms:references (Dublin Core Metadata Initiative, 2018). Here, the prefix "dcterms" refers to the xml namespace of Dublin Core terms http://purl.org/dc/terms/.

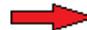

**Figure 1**: showing the example of an ontology reference (indicated with an arrow) added in the standard reference list of a publication. The content and style of the reference are prepared based on the ontology reference template proposed in this work in section 5.

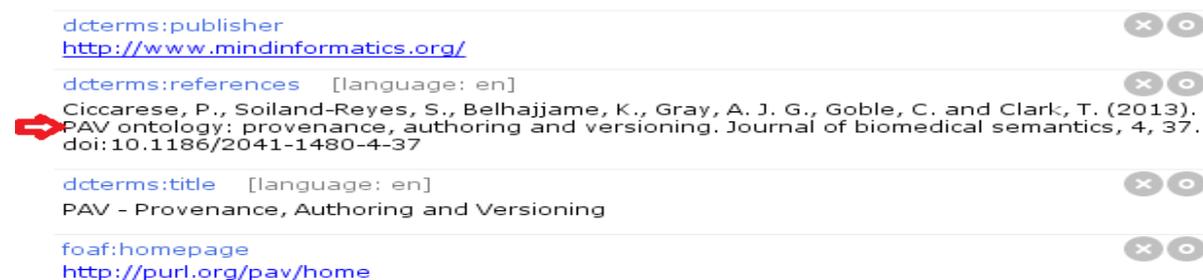

**Figure 2**: showing a citation referring to the journal publication on PAV ontology. The display has been created using the Protégé desktop ontology editing tool (Stanford Center for Biomedical Informatics Research, 2018)

## 5. Content of an ontology reference: our proposal

To derive the content of an ontology reference, first, we have tried to identify and select the fundamental elements for an ontology description and access. For this purpose, we have gone through several numbers of ontology metadata downloaded from the ontology repositories and the general Web. In selecting the elements for ontology reference, we tried to be minimal but yet sufficient to match the motivation of ontology citation as illustrated in section 3. The goal was to keep the reference short and crisp. Table 1 provides a template for referring an ontology. The elements are ordered following the generic pattern author, date, title, source, similar to the APA style usually followed in referring to the traditional publications.

However, the ordering of elements may vary depending upon the publication where the ontology is to be cited. Because each publisher and also depending upon the disciplines, the choice of citation styles varies (see section 1 for details). Table 2 provides an example of an ontology reference created based on the proposed template of Table 1.

**Table 1**. The proposed template for referring an ontology

> Creator. (Date). *Acronym: full name of the ontology*. Version(revision), URI [Ontology file format]

Where,
- *Creator* refers to the main authors of an ontology (and not the contributors). In the case of ontology, it is often the case that besides the creators, there are people who contribute to the ontology in various forms (e.g., participate in the meetings, gives feedback, evaluate the ontology, point to the possible bugs in an ontology, create/assist in preparing the ontology documentation, etc.). We suggest mentioning only the author names. All the author names should be provided. It is suggested to render the author name(s) in the form of *surname, first name*. In case of unavailability of any personal author name, provide the group name, i.e. the name of the organization (or institution) produced the ontology.
- The *Date* is the date of publication (release date) of an ontology. It is suggested to provide the complete date of publication of a particular version of an ontology. The suggested date format is YYYY-MM-DD.
- The *name* of an ontology. The name of an ontology may consist of two parts: *acronym* and *full name* of an ontology. In the case of ontology, the use of the acronym is very common. Most of the ontologies are known by their acronym (e.g., SUMO for Suggested Upper Merged Ontology, DOLCE for Descriptive Ontology for Linguistic and Cognitive Engineering, BFO for Basic Formal Ontology, MOD for Metadata for Ontology Description and publication). Following the acronym, provide the *full name* of the ontology. As indicated above, the acronym and the full name of an ontology should be separated by a colon ":".
- *Version(revision)* is an important information for providing access to a particular version of an ontology. An ontology goes through several revisions and editions, time to time gets updated. As indicated, the version number should be mentioned first and following this, the revision number (if any) of an ontology should be provided within a circular bracket "()".
- *URI*, the *URI* of an ontology, preferably the URI pointing to an ontology file. However, in case of availability of an ontology in multiple file formats, it is suggested to provide the base URI of the ontology.
- *Ontology file format*. Provide the file formats within the square bracket "[]". Also, use commas "," to separate the multiple file formats (e.g., [rdf/xml, owl/xml, obo, n3]).

Example:

**Table 2**. Showing the reference for PAV ontology

> Ciccarese, P. and Soiland-Reyes, S. (2014-08-28). *PAV: Provenance, Authoring and Versioning*. 2.3.1. http://purl.org/pav/ [rdf/xml]

Dutta, B. (2018). Theoretical analysis and propositions for "ontology citation." In Proc. of the Int. Conf. on Exploring the Horizons of Library and Information Sciences: From Libraries to Knowledge Hubs, 7-9 August, 2018 Bangalore, India, pp. 451-458. ISBN 978-93-5311-726-9.

## 6. Conclusion

In the current work, we have investigated the present state of ontology citation and the issues from the theory and practice point of view. It is found that there is no standard practice exist today for ontology citation. People practice differently, some prefer to cite the publication that is based on an ontology, whereas some prefer to cite the ontology link mostly as a footnote in a publication. Both of these practices have several issues as discussed in the paper. In this work, we have advocated citing the source of the ontology providing the complete bibliographic details. Towards this, we have proposed a template mentioning the necessary elements for referring to the source of an ontology. We have illustrated why ontology citation is needed, also provided a set of principles for ontology citation and practices. The main limitation of this work is the proposed citation approach and the reference template are not evaluated. In future, we would be interested to evaluate it by the ontology community. There are many aspects of ontology citation, for instance, the citation index, ontology network analysis, ontology impact factor, including the others as indicated in this paper can be investigated by the researchers interested in the area.